\documentstyle[sprocl]{article}

\def\ra{\rightarrow}

\def\be{\begin{eqnarray}}
\def\ee{\end{eqnarray}}
\def\bq{\begin{equation}}
\def\eq{\end{equation}}
\def\ben{\begin{enumerate}}\def\een{\end{enumerate}}
\def\d {\partial}
\def\L{{\cal L}}
\def\mn{m_N}
\def\mnstar{m_N^\star}\def\Mnstar{M_N^\star}\def\kF{k_F}
\def\mpi{m_\pi}
\def\F1{F_1}
\def\ft1{{\tilde{F}_1}}
\def\ft1omega{\tilde{F}_1^\omega}

\def\prl {Phys. Rev. Lett.}\def\pr{Phys. Rev.}
\def\np{Nucl. Phys.}\def\pl{Phys. Lett.}
\def\la{\langle}\def\ra{\rangle}
\def\O{{\cal O}}
\def\roughly#1{\mathrel{\raise.3ex\hbox{$#1$\kern-.75em%
\lower1ex\hbox{$\sim$}}}}

\def\gsim{\roughly>}
\def\f{{f_\pi^\star}}
\def\barnu{\bar{\nu}}
\def\mec{\epsilon_{MEC}}
\begin{document}
    \title{Chiral Symmetry, Landau Fermi Liquid\\And Dense Hadronic
     Matter~\footnote{\small Talk given at the Joint Session of
97 Korean Physical Society Meeting -- APCTP Workshop on Astro-Hadron Physics,
Seoul, Korea, 25-31 October 1997}}
    \author{MANNQUE RHO}
    \address{Service de Physique Th\'eorique, CEA Saclay,\\
     F-91191 Gif-sur-Yvette, France\\E-mail: rho@spht.saclay.cea.fr}
    \maketitle
  \begin{abstract}
  I describe how effective chiral Lagrangian field theories work
  for nuclei and nuclear matter by going from the deuteron to dense hadronic
matter via BR scaling and
Landau Fermi-liquid fixed point parameters, with possible applications
to relativistic heavy-ion processes and compact ``nuclear stars.''
  \end{abstract}

\section{Introduction}

In this article, I would like to describe in a rather qualitative manner
the recent development in implementing the general framework of effective
field theories in nuclear physics. Some of the matters treated here appeared
in a previous note written for a different occasion \cite{aps}.

It is now generally believed that all
viable theories in all fields of physics --  nuclear, particle and condensed
matter physics, perhaps including string theory
that was formerly considered to be TOE (``theory of everything'') --
are effective field theories.
Indeed,
in a recent {\it Nature} article, Weinberg, in summarizing the present
understanding of what is meant by ``elementary particles" and field theory,
wrote \cite{weinnature}:
{\it We have come to understand that particles may be described at sufficiently
low energies by fields appearing in so-called effective quantum field theories,
whether or not these particles are truly elementary. For instance,
even though nucleon
and pion fields do not appear in the Standard Model, we can calculate the
rates for processes involving low-energy pions and nucleons by using an
effective quantum field theory of pion and nucleon fields rather than of
quark and gluon fields. .... When we use a field theory in this way, we
are simply invoking the general principles of relativistic quantum theories,
together with any relevant symmetries; we are not making any assumption about
the fundamental structure of physics.}

In this article, I illustrate how well
this statement of Weinberg can be applied
to nuclear physics, ranging from the simplest nucleus--  the deuteron --
to the biggest nucleus -- the ``nuclear star.''
In going from a dilute system to
dense matter, I will develop arguments to enable us to go from effective field
theories for the proton to effective field theories for dense matter via Landau
Fermi liquid fixed point theory.

\section{The Lightest Nucleus: Deuteron}
\subsection{Radiative np capture}

Consider the classic case of a process involving two-nucleon systems studied
for almost half a century,
\be
n+p\rightarrow d +\gamma \label{np}
\ee
at thermal energy, with the relative momentum in the center of mass system
$p\simeq 3.4451\times 10^{-5}$ MeV. This process was explained within
10\% accuracy by Austern~\cite{austern} already in 1953
and the remaining 10\% discrepancy
was explained in terms of meson-exchange currents by Riska and
Brown~\cite{riskabrown} in 1974.
I will now describe how one can completely understand this process in
an effective chiral Lagrangian formalism~\cite{pmr}, a truly remarkable
feat for an effective field theory.

For this process, we can start
with a theory defined in the vacuum, that is to say, in  matter-free
space since the two-body system is a dilute one. From the
QCD point of view, the essential physics is dictated by the quark condensate
in the vacuum $\la \bar{q}q\ra$
since apart from the small up and down quark masses, the masses and
couplings of the relevant degrees of freedom, i.e., light-quark hadrons, are dictated by the spontaneously broken
chiral symmetry.
This is because the length scale is set by the quark condensate
in the vacuum. Thus the relevant theory is the one that one should be able
to write down in matter-free space.

Now following the general strategy of
effective field theories, we first have to identify the relevant degrees
of freedom that we would like to treat explicitly and put all irrelevant
degrees of freedom into the constants appearing in the Lagrangian.
Since the process involved (\ref{np}) is a very low-energy process,
we take explicitly as relevant degrees of freedom
the proton and neutron fields denoted as a doublet $N$
together with the
Goldstone excitations of spontaneously broken chiral symmetry,
namely, the pions $\pi^i$. For the moment, other heavy degrees of freedom
such as
the baryon resonance $\Delta$, the vector mesons $\rho$ and $\omega$
and the scalar $\sigma$ will be integrated out so that they will not figure
explicitly in the theory. They will of course figure somewhere in the theory
and I will show where later.

The next step is to write down the most general Lagrangian consisting of
the $N$ and $\pi^i$ fields consistent with the general properties Weinberg
is referring to above,  notably, spontaneously
broken chiral symmetry. The Lagrangian will contain, in addition to
bilinears in the $N$ field, terms involving $4N$, $6N$ etc. to all orders
in power of fermion fields suitably coupled to Goldstone pions
and since light-quark masses are not zero, though small, there should be
terms involving quark mass terms. In nuclear physics at low energies,
the nucleon can be considered as heavy. When the nucleon is treated
in that way,
the leading Lagrangian, when expanded, can be written as
\be
\L&=& N^\dagger \left(i\d^0 +\frac{\nabla^2}{2M}\right)N -\frac{g}{f_\pi}
\nabla {\bf \pi}\cdot N^\dagger{\bf \tau} \sigma N +\frac 12 (\d^0{\bf \pi})^2 -
\frac 12(\nabla{\bf \pi})^2 -\frac 12 m_\pi^2 ({\bf \pi})^2 \nonumber\\
&& + \frac{C_1}{f_\pi^2} (N^\dagger N)^2 +\frac{C_2}{f_\pi^2} (N^\dagger\sigma
N)^2 +\cdots \label{L}
\ee
where ${\bf \pi}$ is the triplet pion field.
Counting rules can be devised in such a way that one can
do a systematic expansion
in some momentum scale $Q$ being probed that is in some sense small compared
to the typical chiral scale $\Lambda$ which can be taken to be roughly
of the mass of the heavy particles that have been
integrated out. If this is effectuated to
all orders, then according to the axiom given by Weinberg~\cite{weinnature},
we are in principle doing a full theory.

Let us see what this scheme means for nuclear interactions. In so doing
I will uncover what is called ``chiral filter phenomenon" in nuclear
processes. A physical amplitude involving $E_N$ external nucleon lines
can be written as
\be
A\sim Q^\nu F(Q/\Lambda)
\ee
where $F$ is a slowly varying function of the dimensionless quantity
$Q/\Lambda$. Given that $Q$ is small compared with the chiral scale,
the idea is to calculate to the highest order possible
in $\nu$ and sum the terms to
the order calculated. If one can do this to all orders, as mentioned,
then one is doing
the full theory. With the chiral Lagrangian that we are concerned with,
the counting rule can be readily deduced by looking at the Feynman diagrams.
One finds
\be
\nu=4-2C+2L-(E_N/2+ E_{ext}) + \sum_i V_i \bar{\nu}_i\label{index}
\ee
where $C$ stands for the number of clusters, $E_N$ the number of
external incoming and outgoing nucleon lines, $E_{ext}$ the number of
external fields, $L$ the number of loops and $V_i$ the number of vertices
of type $i$ and
\be
\bar{\nu}_i= d_i +\frac{n_i}{2} +e_i -2\label{subindex}
\ee
with $d_i$ the number of derivatives, $n_i$ the number of nucleon lines
and $e_i$ the number of external fields entering the $i$th vertex.

In this paper we will be concerned with at most one slowly varying external
field, so we will have $E_{ext}=1$. Obviously
 $e_i=1$ will appear only once if at all.
For two-body exchange currents, we are concerned with an irreducible graph
with $E_N=4$ and $C=1$. So the important quantities in (\ref{index}) are
the number of loops $L$ and the $\barnu$. For a given $L$, therefore,
only $\barnu$ matters.
The particular structure of chiral symmetry requires that
\be
\bar{\nu}_i\geq 0.
\ee

We can now state the ``chiral filter phenomenon" which will figure prominently
in what follows.
In the form it was first stated~\cite{kdr}, the general argument
that follows from effective chiral Lagrangians was not used. What the statement
says is that while nuclear forces involve long-range and
short-range interactions on the same footing and hence have to be taken into
account at the same time, the response to a slowly varying electroweak
field {\it screens} short-distance physics, thereby causing
the effect of soft-pion mediated
process to show up prominently (unless accidentally suppressed by kinematics).
One can see this to the lowest order in the chiral
counting. Take the two-nucleon potential generated by one-pion exchange
and the one generated by a contact four-Fermi interaction in Eq.~(\ref{L}).
The one-pion exchange involves two vertices each of which has
index $\bar{\nu}_i=0$ since there is one derivative in the pion-nucleon
coupling and two nucleons attached to the vertex, $n_i=2$.
Higher-order terms in derivative will bring in higher
power and be suppressed.
So at the zero-loop level, it is this tree term that matters.
But the same is true with the four-Fermi interaction with zero loops
as there is no derivative at the vertex but four nucleons enter with $n_i=4$,
so again $\barnu=0$. The one-pion-exchange potential is
the longest-range one in two-nucleon systems. Now four-Fermi interactions
represent the short-range part of the potential in which heavy mesons
representing the degrees of freedom lying above the chiral scale $\Lambda$ have
been integrated out. Thus we conclude that as far as the chiral counting
is concerned both the {\it longest-range} and
{\it shortest-range} potentials
contribute on an equal footing and cannot be
separated.

Consider attaching an external field to the same two-body system.
Attaching an external field to the one-pion-exchange term does not modify
the index $\barnu$ since $e_i+d_i=1$, hence we still have $\barnu=0$ in the
leading order but the four-Fermi interaction term requires one derivative
in addition to $e_i=1$ leading to an index
$\barnu\geq 2$. Therefore in contrast to the nucleon-nucleon potential,
short-range interaction terms are naturally suppressed relative to the
one-pion-exchange term
when attached to an external field~\cite{mr91}. This is true for a slowly
varying electroweak current in general.

This ``chiral filtering" is both a good news and a bad news. It is
a good news in that meson-exchange currents can be under control
with the dominance of Goldstone pions without interference from
poorly-understood short-range degrees of freedom. It
is a bad news since the pion dominance means that unless accidentally
suppressed, pions will not allow us to learn short-distance physics
through exchange currents. To some who would like to see a ``smoking
gun" of quark-gluon degrees of freedom in nuclear processes, this may be
disappointing.


I should explain (in more detail than on other topics)
 what went into the calculation of the cross section for (\ref{np}).
Imagine that we have the process occurring between a proton and a
neutron interacting to all orders through the chiral Lagrangian of
the form (\ref{L}). The initial state is the scattering state in
$^1S_0$ and the final state is the bound (deuteron) $^3S_1$ state
with a small D-state admixture. The electromagnetic current --
which is predominantly magnetic dipole ($M1$) -- connects the
initial and final states. Both the bound state and the scattering
state with a large scattering length (with $a_s=-23.75$ fm) are not
amenable to a straightforward chiral perturbation expansion because
of infrared divergences and so a different strategy is needed as
discussed in~\cite{PKMR}. For the moment, we shall bypass this
difficulty  by observing that what we are interested in is the
meson-exchange current contribution {\it relative} to the
single-particle matrix element for which we can take the most
realistic wave functions available for the initial and final
states. Such wave functions are indeed in the market~\footnote{As
discussed by Tae-Sun Park in this meeting, the wave function has
actually been calculated from ``first principles." This will be
briefly described below and will be published in a greater
detail~\cite{PKMR}.}, namely those computed from the Argonne
$v_{18}$ potential~\cite{v18}. This procedure is not exactly a
systematic chiral expansion since the infrared-divergent reducible
diagrams are summed to all orders (in the form of solving the
Schr\"odinger equation) while irreducible graphs are computed to
the NNL order only. However the NNL order essentially saturates the
irreducible graphs within the uncertainty associated with the
short-distance part of the wave function, which is of the order of
less than 1\%, so the scheme is consistent as far as this
calculation is concerned.

To the extent that the wave functions are very accurate, the single-particle
matrix element will also be. One can gauge this by looking at the
prediction for the $^1S_0$ scattering length and the static properties of
the deuteron, all of which are remarkably accurately given
by the Argonne $v_{18}$ potential.
It is therefore convenient to look only at the exchange-current
corrections {\it relative} to the single-particle matrix element. The dominant
one-pion exchange matrix element is ${\O} (Q^2)$ relative to the single-particle
matrix element. One-loop radiative corrections are further suppressed
by the same order.
The four-Fermi interactions found to be suppressed by the ``chiral filtering"
enter at the range of uncertainty associated with the short-range correlation
in the wave functions which cannot be accessed by chiral perturbation
theory. In fact, the suppression of the zero-range operators due
to the correlation function represents in an indirect way this part of
physics~\cite{PKMR,lepage}.

Now calculated
with the single-particle matrix element alone, the cross section comes out
to be
\be
\sigma_{imp}=305.6\ \ {\rm mb}
\ee
which differs by about 9\% from the experimental value
\be
\sigma_{exp}=334.2\pm 0.5\ \ {\rm mb}.
\ee
The chiral Lagrangian treatment, taking into account the short-range
uncertainty mentioned above, gives an accurate result, accounting fully
for the missing 9\%,
\be
\sigma_{theory}=334\pm 3\ \ {\rm mb}
\ee
where the quoted error represents theoretical uncertainty associated with
short-distance physics.

Perhaps much less solid theoretically but more spectacular is the
electrodisintegration
of the deuteron, a sort of inverse process to the np capture,
\be
e + d \rightarrow e+n+p.
\ee
If one applies the same formalism as in the np capture, it is found that
the meson-exchange current effect, while small in the np capture, becomes
big because of a substantial cancelation at finite momentum transfer
between the S-state and D-state components
of the deuteron wave functions. The chiral expansion to the next-to-next order
described above turns out to work well up~\cite{frois} to a large
momentum transfer of order $\sim 1$ GeV.
While it has not been checked in detail that other corrections
remain negligible here, the important
presence of the mesonic current is clearly
exhibited in this process. It remains to prove that higher order
chiral corrections are indeed suppressed at large momentum transfers involved.
\subsection{Deuteron properties in effective field theories}

In the above calculation, we took simply the best deuteron wave
function available in the literature, bypassing actually working
out the relevant chiral perturbation scheme. Here I summarize how
this calculation can be done with a great accuracy within the
framework of an effective field theory~\cite{PKMR}.

\begin{table}
\caption[deuteronTable]{Deuteron properties and
the M1 transition amplitude entering into the $np$ capture for
various values of the cut-off $\Lambda$.}\label{table1}
\vspace{0.2cm}
\begin{center}
\footnotesize
\begin{tabular}{ccccccc}
$\Lambda$ (MeV) & $150$ & $198.8$ & $216.1$ & $250$ & Exp. & $v_{18}$\cite{v18} \\
\hline
$B_d$ (MeV) & $1.799$ & $2.114$ & $2.211$ & $2.389$ & $2.224$ & 2.224\\
$A_s$ ($\mbox{fm}^{-\frac12}$)
   & 0.869 & 0.877 & 0.878 & 0.878 & 0.8846(8) & 0.885 \\
$r_d$ (fm)
   & 1.951 & 1.960 & 1.963 & 1.969 & 1.966(7) & 1.967 \\
$Q_d$ ($\mbox{fm}^2$)
   & 0.231 & 0.277 & 0.288 & 0.305 & 0.286 & 0.270 \\
$P_D$ (\%)
   & 2.11 & 4.61 & 5.89 & 9.09 & $-$ & 5.76 \\
$\mu_d$
   & 0.868 & 0.854 & 0.846 & 0.828 & 0.8574 & 0.847 \\
$M_{\rm 1B}$ (fm)
   & 4.06 & 4.01 & 4.00 & 3.97 & $-$ & 3.98
\end{tabular}
\end{center}
\end{table}

The process involved is a very low-energy process, with the energy
scale probed much less than the pion mass $\sim 140$ MeV. So in
calculating the single-particle $M1$ matrix element, we can
integrate out {\it even} the pion field as well. Thus the effective
Lagrangian we are left with consists then of terms bilinear,
quartic etc. in the nucleon field with the coefficients of the
higher-Fermi field terms to be determined from experiments. The
higher-Fermi field terms are local and hence are delta functions
and derivatives of delta functions in coordinate space that require
regularization. In \cite{PKMR}, the theory is done up to the
next-to-leading order in the chiral counting. With the nucleon
matter fields (considered as heavy-fermion fields) 
alone there are no ``irreducible loops" (if pions are
present, then there will be) and ``reducible loops" are summed to
all orders corresponding to solving the Lippman-Schwinger equation.
There are then five constants that appear in the potential in the
theory: $C_{0,2}$ for spin $S=0,\ 1$ and
$D_2$ for $S=1$ in the form
\be
V({\bf q}) = \frac{4\pi}{M} \left(C_0 + (C_2 \delta^{ij} + D_2
\sigma^{ij}) q^i q^j \right)
\label{Vq}\ee
where $M$ is the nucleon mass and
\be
\sigma^{ij}=\frac{3}{\sqrt{8}}\left(\frac 12 (\sigma_1^i\sigma_2^j+
\sigma_1^j\sigma_2^i)-
\frac 13 \delta^{ij}\sigma_1\cdot\sigma_2\right).
\ee
These constants should in principle be calculable from the
fundamental QCD Lagrangian but nobody knows how to do such
calculations at the moment and it will be some time before we will
see any results. The strategy at our disposal is to fix them from
experiments. This can be done accurately in the present case. The
four constants $C_{0,2}$ for the two spin channels can be fixed from 
the scattering lengths and
effective ranges for the scattering in the $^1S_0$ and $^3S_1$ channels 
and $D_2$
can be fixed by the deuteron $D/S$ ratio. The theory will be
regularized with a cut-off $\Lambda$ which should be of order
$m_\pi$, the lightest degree of freedom integrated out. As is now
well-known, the dimensional regularization, simpler though it may
be, is not very useful, giving a completely wrong answer if naively
applied. Using the cut-off as regularization is very natural in
effective theories as discussed in \cite{shankar}. One should not
take the cut-off too high or too low and the optimal value found in
\cite{PKMR} was $\Lambda\sim 200$ MeV.

One finds that with this cut-off, the phase-shift in the $^1S_0$
channel is accurately described up to the center-of-mass momentum
$\sim 70$ MeV and the deuteron properties and the $M1$ matrix
element come out in beautiful agreement with experiments as one can
see in Table 1. For $\Lambda= 216.1$ MeV, the results are quite
spectacular~\footnote{It should be underlined here that the quadrupole moment
comes out correctly in this theory while as the authors of \cite{v18} admit, 
it does not with the most accurate phenomenological approach with the
Argonne $v_{18}$ potential. It would be important to understand why this
is so.}. Furthermore the insensitivity to the precise value of
the cut-off value can be taken to be an evidence that the effective
field theory does work well here.

The cut-off used in \cite{PKMR} is of a Gaussian type which brings
in terms higher order than the next-to-leading order used in the
potential (\ref{Vq}). To be consistent with the power counting, one
would have to incorporate corresponding ``counter terms'' in
$V({\bf q})$ -- since there are no irreducible loops in the theory,
these are the only higher order terms appearing in the expansion --
although the good agreement indicates that the latter must be
insignificant. As one goes up in energy, the pion field would have
to be considered explicitly. The introduction of the pion as well
as doing higher order calculations -- including pion loops -- would
teach us how to test the interplay between the breakdown of an
effective field theory and the emergence of a ``new physics,'' an
issue currently figuring importantly in particle physics in the
effort to go beyond the Standard Model.

\section{BR Scaling}
\subsection{Nuclear axial-charge transitions}

Another place where the chiral filtering is visibly operative and
where ``new physics" could potentially enter as the matter becomes
dense is in axial charge transitions in heavy nuclei, i.e,
\be
J^+\leftrightarrow J^- \ \ \ \Delta T=1.
\ee
If $J=0$, this process is analogous to a pion decaying into the vacuum
carrying an interesting information on the ``vacuum property" of
nuclear ground state. Warburton~\cite{warburton}
studied extensively  this class of
transitions in light, medium and heavy nuclei, and obtained an important result that
as one goes to heavier nuclei the effect of the pion exchange
in the axial charge matrix elements becomes stronger. He defined
a quantity called $\epsilon_{MEC}$
\be
\mec= \frac{M_{exp}}{M_{sp}}
\ee
where $M_{exp}$ is the experimentally measured matrix element of the
axial charge operator and $M_{sp}$ is the theoretical matrix element
of the single-particle axial charge operator calculated with the
best possible nuclear wave functions available in the literature.
Since this quantity involves both
experimental and theoretical quantities, it is not quite
what one would call experimental value. There is an inherent uncertainty associated with
the single-particle matrix element.
What is however quite significant in the study of
Warburton is that unlike in the case of electromagnetic exchange currents
the effect of the chiral-filtered pions can be enormous. Indeed in light
nuclei, $\mec$ is around 1.5, that is, the exchange correction is 50 \%
in the matrix element. This is a huge correction.
What is more
significant is that in heavy nuclei, the effect is even more dramatic.
In lead region, Warburton found
\be
{\mec}=1.8\sim 2.0.\label{warburton}
\ee
The range is the uncertainty involved in the theoretical single-particle
matrix element alluded to above.

There is a simple way of getting the enhancement (\ref{warburton}). This
can be done by
combining the chiral filter mechanism together with
what is known as ``BR scaling"
in dense medium which I shall now explain. The idea involves once more
the general philosophy of effective Lagrangians but extrapolated further
into the regime where matter is dense and where direct measurements are
not readily available.

If physics does not change drastically from light to heavy nuclei, one
may start with a Lagrangian like (\ref{L}) and then compute systematically
the effect of the medium by suitably accounting for additional scales
brought in by matter density. There are efforts to do this sort of
calculations. Here I will consider approaching from the other extreme (say,
a ``top-down" approach)
where possible nonperturbative effects associated with the medium are taken
into account {\it ab initio}
in a manner consistent with the notion of chiral effective theories.
This has an advantage in that physics under extreme conditions such as
the state of dense matter encountered in compact neutron stars and relativistic
heavy-ion collisions can be treated on the same footing. Viewed in this way,
calculating the enhancement (\ref{warburton}) will be a low-order calculation
whereas starting from (\ref{L}) would require ``high-order" calculations.

In writing down Eq.~(\ref{L}), I emphasized that it has particles whose
parameters are defined in the absence of background matter. Now consider
a particle, a fermion or a boson, propagating in a medium consisting of
matter in a bound state like in the interior of a very heavy nucleus.
For this, I can start with a Skyrme type Lagrangian containing only meson
fields. Imagine having a realistic Lagrangian of such type containing not
only pions as in the original Skyrme Lagrangian but also vector mesons
and other heavy mesons. A nucleon with this Lagrangian comes out as a
soliton, ``skyrmion," with baryon number $B=1$ which is just the winding
number of the soliton. The same Lagrangian in principle
can describe the deuteron, triton
... and $B=\infty$ nucleus, all arising from the same Lagrangian. At present
we do not know how to write such a Lagrangian and hence we do not know
how to compute, for instance, the binding energy,
the equilibrium density of nuclear
matter and nuclear matrix elements of currents. What is known is that
the deuteron and nuclear force can be reasonably understood even
from a drastically simplified skyrmion Lagrangian~\cite{CND}.

Given a realistic chiral Lagrangian of the skyrmion type, the question one can
ask is: How does a hadron propagate in a medium defined by a density
$\rho$, say ? The most obvious thing to do is then to write an effective
Lagrangian that has all the right symmetries of the original theory, QCD,
but suitable in the background defined in the presence of a medium.
In QCD, the quantity that reflects the background or the ``generalized vacuum"
is the quark condensate
and since the background is changed, we expect that the condensate would be
suitably changed. Let me denote the modified condensate by putting an asterisk
\be
\la \bar{q}q\ra^\star\neq \la\bar{q}q\ra_0, \ \ \ \rho\neq 0.
\ee
Since the condensate is modified, all the associated quantities such
as light-quark hadron masses, the pion decay constant $f_\pi$ etc. will
be suitably modified. I will denote them with an asterisk on top. By following
the strategy of preserving the same symmetry present in matter-free
space {\it except that asterisked parameters enter}, it is possible to
establish the scaling~\cite{BRscaling}
\be
m_V^\star/m_V\approx M_N^\star/m_N\approx m_\sigma^\star/m_\sigma\approx
f_\pi^\star/f_\pi\approx (\la \bar{q}q\ra^\star/ \la\bar{q}q\ra_0)^n\label{BR}
\ee
where the subscripts $V$, $\sigma$, and $N$ stand, respectively, for
(light-quark) vector
meson, scalar meson and nucleon fields and the index $n$ is some power that
depends on specific models (for the simplest
Skyrme model, $n=1$, for the NJL model,
$n=1/2$ etc.). An effective Lagrangian of the type (\ref{L}) with its
parameters given by (\ref{BR}) can then describe, at tree order, fluctuations
around the state defined by density $\rho$.

At present there is no systematic derivation of such an effective Lagrangian
from first-principle arguments. As such, the scaling (\ref{BR}) is not a
relation that can be used in {\it any}
Lagrangian field theory dealing with nuclear matter.
It should be considered as a particular parameterization with a given Lagrangian
of the type I have been considering. Thus the quantities with such scaling can
have meaning only as {\it parameters} of a specific theory and it would be too hasty to identify them as ``physical'' masses and constants. The only
quantity that is physically meaningful is the measurable one.

One way to ``derive'' the scaled Lagrangian is to look for a non-topological
soliton of the effective action arising from a high order chiral perturbation
theory. As suggested by Lynn~\cite{lynn}, it could be a ``chiral liquid'' that defines
the Fermi sea with a given Fermi momentum $k_F$. One can
identify this as a ``chiral-scale'' decimation in the renormalization
group approach mentioned below, with the cutoff set at the chiral scale
$\Lambda \sim \Lambda_\chi \sim 1$ GeV. Once such a ``chiral liquid'' state is
obtained, then the scaling parameters will be defined in fluctuations around
the chiral liquid state in what I would call ``Fermi-liquid scale
decimation~\cite{SBMR}.'' I will return to this matter.

Let us now go back to Warburton's $\mec$ in heavy nuclei for which we will
take $\rho\approx\rho_0$. Suitably coupling the axial current to
a BR effective Lagrangian, one can calculate and find~\cite{pmraxial}
\be
\mec={\Phi}^{-1} (1+R)\label{mecth}
\ee
where
\be
\Phi:=f_\pi^\star/f_\pi\approx m_V^\star/m_V \cdots
\ee
and $R$ is the ratio of the matrix elements of the meson-exchange
axial charge operator over the single-particle axial charge operator.
The meson-exchange operators are given in chiral perturbation theory
to the next-to-next-to leading order in the chiral expansion as in the
electromagnetic case, again dominated by the pions due to the chiral filter
as explained in~\cite{kdr}. The ratio $R$ does not depend much on how one
calculates the matrix elements, that is, nuclear model-independent,
and depends only slightly on density. For heavy nuclei, it comes out
to be $R\approx 0.5\sim 0.6$. The quantity we need to compute $\mec$
is $\Phi$, the only quantity that knows that nuclear matter ``vacuum"
is different from the matter-free vacuum. There are two ways known to
get this quantity -- and this is not given by the strategy of effective chiral
Lagrangian field theory. One is to use the Gell-Mann-Oakes-Renner formula
for the pion embedded in nuclear medium, the other is to do a QCD sum-rule
calculation for the vector-meson mass in medium. While both quantities in medium are
not without ambiguity, they nonetheless give
the same answer. The result by the latter method
at $\rho=\rho_0$ is~\cite{jin}
\be
m_V^\star/m_V=\Phi (\rho)=0.78\pm 0.08, \ \ \rho=\rho_0.\label{Phi}
\ee
With this value, (\ref{mecth}) gives
\be
\mec=1.9\sim 2.0.
\ee
This agrees with Warburton's analyses~\footnote{One could
of course calculate corrections to the chiral-filtered pionic
contribution without
invoking BR scaling but instead using a vacuum-defined chiral Lagrangian and
explicitly incorporating other degrees of freedom (such as an effective scalar
meson $\sigma$ and light-quark vector mesons) and get the required
enhancement~\cite{riska}. The two methods must be equivalent to leading order
at nuclear matter density.}.
\subsection{Kaon production and kaon condensation}

It is easy to generalize the formalism to $SU(3)$ flavors and study
fluctuations in the strangeness directions. For instance, one could
look at the production of kaons in dense medium in heavy-ion collisions.
Once the ground state is defined in terms of BR scaling chiral Lagrangians,
fluctuations are then automatic at tree level, combining both flavor
$SU(3)$ symmetry and chiral symmetry. How this can be done is discussed
in~\cite{SBMR}. Some of the predictions made in this way have been tested
by experiments recently performed at GSI (e.g., FOPI and KaoS) and
are fairly well
confirmed~\cite{llb97,LLB2}.
Extended smoothly beyond nuclear matter density, the theory can make
predictions on possible phase transition with condensation of kaons
at a density $\rho\sim 3\rho_0$ in compact-star matter like in nucleon
stars with a fascinating consequence on the formation of small
black holes and on
the maximum mass of stable neutron stars etc. \cite{brownbethe,LLB2}.
\section{Nuclear Matter as a Fermi Liquid}
\subsection{Fermi liquid as a renormalization-group fixed point}

Up to this point, I have not discussed how the ground state, namely
nuclear matter, comes out in this description. I shall now ``map" the
chiral Lagrangian with BR scaling treated in mean field to Landau
Fermi-liquid theory of nuclear matter developed by Migdal~\cite{migdal}.
The idea is based on two observations. The first is that relativistic
mean-field theory for nuclear matter is known to be interpretable as
equivalent to Landau Fermi-liquid theory. For instance, Walecka's mean
field theory has been shown to be one such theory~\cite{matsui}. The second
is that Landau Fermi-liquid theory is a renormalization-group fixed point
theory \cite{shankar}. This observation allows one to formulate a many-body
problem from the point of view of effective field theories
which is clearly what is needed
to go further into the unknown regime of high density.
 In a recent paper, Brown and I \cite{br96} argued that
the BR scaled chiral Lagrangian in a simplified form, when treated at
the mean field level, is equivalent to a Walecka-type mean field theory.
It is therefore quite logical that the BR scaled chiral Lagrangian
mean-field theory is equivalent to Landau Fermi-liquid fixed point theory.
The relevant arguments linking various elements of the theory are found
in~\cite{SBMR}.

The crucial link is found at the stage of the second -- ``Fermi-liquid'' --
decimation that integrates out excitations of the scale $\tilde{\Lambda}$ around the Fermi surface defined by the Fermi momentum $\kF$ and then does the rescaling.
The main ingredient is the renormalization group-flow result that
there are two fixed-point quantities in the theory~\cite{shankar}.
One of them is the effective
mass of the nucleon $m_N^\star$ and the other is the Landau interaction
${\cal F}$.
By Galilean invariance, the effective mass is related to the $l=1$
Landau parameter $F_1$ as
\be
\mnstar/\mn=1+\frac 13 \F1=\left(1-\frac 13 \tilde{F}_1\right)^{-1}
\label{landaumass}
\ee
where $\tilde{F}_1:=(\mn/\mnstar)\F1$.
Now using that the Walecka model is equivalent
to Landau Fermi liquid, we deduce that $\tilde{F}_1$
gets a contribution from the
$\omega$ channel, say, $\ft1omega$. Due to chiral symmetry, there is also the Goldstone pion
contribution through a Fock term to $\tilde{F}_1$
which can be explicitly calculated.
Thus
\be
\tilde{F}_1=\ft1omega +\tilde{F}_1^\pi
\ee
with
\be
\ft1omega=3(1-\mn/\Mnstar)=3(1-\Phi^{-1})\label{ft1omega}
\ee
and
\be
\tilde{F}_1^\pi
=-\frac{9f_{\pi NN}^2 \mn}{8\pi^2 \kF}\left[\frac{\mpi^2+ 2\kF^2}{2\kF^2}
\ln\frac{\mpi^2+4k_F^2}{\mpi^2} -2\right]\label{ft1pi}
\ee
where $\f_{\pi NN}\approx 1$ is the nonrelativistic $\pi N$ coupling constant.
Note that (\ref{ft1pi}) is precisely
determined once the Fermi momentum is given, say, $\approx -0.153$
at normal matter density. The important point here is that the effective mass
gets contributions from the (BR) scaling parameter (\ref{ft1omega}) {\it and}
the pion. The pion comes in as a perturbative correction to the nonperturbative ``vacuum''
contribution given by $\Phi$. The effective (Landau) mass
(\ref{landaumass}) is therefore
\be
\mnstar/\mn=\left(\Phi^{-1}-\frac 13 \tilde{F}_1^\pi\right)^{-1}\label{Lformula}
\ee
which at $\rho=\rho_0$ predicts
\be
\mnstar(\rho_0)/\mn \approx 0.69.\label{effmass}
\ee
This is a genuine prediction which is supported by the orbital gyromagnetic
ratio in heavy nuclei, discussed below.
It is also consistent with the QCD sum rule
calculation of the nucleon mass in medium~\cite{mnsr},
\be
(\mnstar(\rho_0) /\mn)_{QCD} =0.69^{+0.14}_{-0.07}.\label{mnqcd}
\ee

Now having the relation between the fixed point $\mnstar$ and $\Phi$ (plus
the calculable pionic term), we can derive
various interesting and highly nontrivial relations applicable to
 long-wavelength
processes~\cite{FR95}.
For instance, the EM convection current for a nucleon on the Fermi surface
which can be written down on the basis of $U(1)$ gauge invariance can
be derived from our chiral Lagrangian:
\be
{\bf J}=g_l \frac{\bf p}{\mn}
\ee
where $g_l$ is
the orbital gyromagnetic ratio given by
\be
g_l=\frac{1+\tau_3}{2}+\delta g_l
\ee
with
\be
\delta g_l=\frac 49 \left[\Phi^{-1} -1 -\frac 12 \tilde{F}_1^\pi\right]\tau_3.
\label{deltagl}
\ee
I should stress that this relation is highly non-trivial for several
reasons. First of all, the isoscalar current is given by ${\bf J}^{(0)}
={\bf p}/2\mn$, so the scaling mass $\Mnstar$ does not figure in the
current (this is an equivalent to ``Kohn theorem" in condensed matter physics)
and secondly the many-body nature of the system is manifested only in
the isovector part through $\delta g_l$. Given the numerical value (\ref{Phi})
at nuclear matter density, we get
\be
\delta g_l=0.23\tau_3.
\ee
This should be compared with $\delta g_l^{proton}=0.23\pm 0.03$ obtained
from a dipole sum rule~\cite{schumacher} in $^{209}$Bi and with $\delta
g_l^{proton}\approx 0.33$, $\delta g_l^{neutron}\approx -0.22$ obtained
from an analysis of the magnetic moments in the $^{208}$Pb
region~\cite{yamazaki}.

The deviation of the nucleon effective mass from the ``universal" scaling
factor $\Phi$, (\ref{Lformula}), arises from the presence of the Goldstone pions. In the skyrmion description,
the difference arises from the fact that aside from the known
current algebra term, an additional
term -- Skyrme quartic term -- is needed for stabilizing the soliton
that metamorphoses into the physical
nucleon. Expressed in terms of physical variables, the difference can be attributed to
the fact that the axial coupling constant $g_A$ can scale in nuclear
medium~\cite{BRscaling,FR95}.
It turns out that
\be
\frac{\mnstar}{\mn}=\left(\frac{g_A^\star}{g_A}\right)^{\frac 12} \Phi.\label{skyrmion}
\ee
Comparing with (\ref{Lformula}), we find that
\be
\frac{g_A^\star}{g_A}= (1+\frac 13 F_1^\pi)^2
=(1-\frac 13\Phi \tilde{F}_1^\pi)^{-2}.\label{gaskyrmion}
\ee
For nuclear matter ($\rho\approx \rho_0$), this predicts~\footnote{Clearly
the formula (\ref{gaskyrmion}) cannot be valid beyond a certain density
$\gsim \rho_0$. It would be nice to show  that $g_A^\star\approx 1$ is a fixed point.}
\be
{g_A^\star}(\rho_0)\approx 1.
\ee
This is quite close to what is found in nature~\cite{gaexp}. The same result
was obtained many years ago in terms of the Landau-Migdal parameter
$g_0^\prime$ in the $\Delta N$ channel~\cite{ROW} which has recently been
interpreted as a counter term in higher order
chiral expansion~\cite{PJM}. The relationship
between these different interpretations is not yet understood and remains to
be clarified.
\subsection{A BR-scaling Lagrangian model}

This close agreement of the chain of predictions with experiments can be taken to confirm the
validity of the notion that the scaling (\ref{BR}) --
initially introduced as a vacuum change --
is associated with the Fermi liquid fixed point in many-body interactions.
 So far we have been looking at
fluctuations on top of the ground state. What about the ground state itself?
The chain of arguments developed thus far can be encapsulated into
a simple effective Lagrangian of the form~\footnote{This
should be understood in the sense of the effective action in the mean field sense,
with $\delta S^{eff}|_{\omega^\star,\phi^\star,...}=0$ for
$S^{eff}=\int \ d^4 x\ \L_{BR} (x)$. Fields not figuring in the mean field such as
pion field are not explicited here. However once the mean field is defined, fluctuations
into strange and non-strange directions can be described by
restoring pion, kaon,... fields
in a way consistent with chiral symmetry. Phenomenology seems to
require that $g_v^\star/g_v$ scale  whereas no such requirements
exist for the scalar constant $h$.}
\be
\L_{BR} &=& \bar{N}(i\gamma_{\mu}(\d^\mu+ig_v^\star\omega^\mu )-
M_N^\star+h\phi )N
\nonumber\\
& &-\frac 14 F_{\mu\nu}^2 +\frac 12 (\partial_\mu \phi)^2
+\frac{{m^\star_\omega}^2}{2}\omega^2
-\frac{{m^\star_s}^2}{2}\phi^2\label{toyBRPP}
\ee
where we have retained only the $\omega$ field and the effective scalar
field $\phi$ in the meson sector eliminating the pion field from the
chiral Lagrangian since we are to interpret it as an effective one to be
considered only in the mean field. I have written this Lagrangian in analogy to
Walecka's original linear $\sigma\omega$ model but it would be more
appropriate to
consider it as a Lagrangian that obtains in a chirally invariant way from
one with 2-Fermi and 4-Fermi interactions using massive auxiliary fields
$\omega$ and $\phi$~\footnote{Thus the $\phi$ is a chiral singlet instead
of the fourth component of $O(4)$ as in the linear sigma model.}.
Treated in the standard manner as for the Walecka model,
this effective Lagrangian describes nuclear matter fairly accurately.
For instance
with the physical masses, $m_N=939$ MeV, $m_\omega=783$ MeV and $m_s=700$
MeV and the parameters, $h=6.62$, $g_v=15.8$ and assuming the
scaling
\be
\Phi(\rho)=(1+0.28\rho/\rho_0)^{-1}
\ee
normalized so that the known value (\ref{Phi}) is reproduced at $\rho=\rho_0$,
we get the binding energy $B$, the equilibrium Fermi momentum $k_F$
and the compression modulus $K$:
\be
B=16.0\ \ \ {\mbox MeV}, \ \ \ k_F= 257\ \ \ {\mbox MeV}, \ \ \ K=296\ \ \
{\mbox MeV}.
\ee
The corresponding effective mass of the nucleon at the minimum is
\be
\mnstar=\Mnstar-h\la\phi\ra^{\star}=0.62 m_N
\ee
which should be compared with (\ref{effmass}) and (\ref{mnqcd}).
The nuclear matter property so obtained
is quite close to that obtained from an effective chiral
Lagrangian constructed based on {\it naturalness condition}~\cite{furnstahl}.
It has recently been shown that this model is entirely
consistent with thermodynamics~\cite{SMR},
an important point for extrapolating to the regime relevant to relativistic
heavy-ion interactions. In particular, the relativistic Landau Fermi-liquid
formulas~\cite{baymchin}
\be
K&=& \frac{3k_F^2}{E_F} (1+F_0),\\
c_1&=& v_F (\frac{E_F}{3\mu} (1+F_0))^{\frac 12},\\
E_F&=&\mu (1+F_1/3)
\ee
where $K$ is the compression modulus, $E_F=\sqrt{k_F^2 +{m_N^\star}^2}$
and $\mu$ the chemical potential are all satisfied. Being a Lagrangian
field theory, the energy-momentum conservation is automatic.

\subsection{Matter under extreme conditions}

The idea developed here allows one to explore what happens when matter is compressed
to a density greater than normal. This is a relevant issue for on-going
experiments in relativistic heavy-ion collisions and for understanding such compact stars as neutron stars. Suppose one would like to probe
the regime where $\rho > \rho_0$. Instead of approaching this regime
with a Lagrangian defined at $\rho=0$ as is done conventionally, I would
like to consider fluctuations
around $\rho\approx \rho_0$ with the effective Lagrangian
{\it defined} at that point. The advantage in doing this is that even if fluctuating around
the $\rho=0$ vacuum were a strong-coupling process and hence required a high-order calculation, fluctuations around the ground state at $\rho=\rho_0$ could
be weak-coupling allowing for a tree-order or at most a next-to-leading calculation.
Indeed the recent elegantly simple explanation of
the CERES dilepton data~\cite{ceres} by Li, Ko and Brown (LKB)~\cite{LKB}
is a nice example of such an application. Here one is probing hadronic matter
at a density $\sim 3\rho_0$ at some high temperature. In the
LKB approach, the dileptons measured in the experiments are interpreted as
arising from mesons in a heat bath with their masses scaled as
(\ref{BR}). The result is consistent with a quasiparticle picture
for both nucleons and mesons in a heat bath.

The argument developed to link effective chiral Lagrangians and Fermi-liquid theory
is manifestly tailored for very low-energy
excitations for which Landau quasiparticle picture is valid. For instance,
in describing nuclear matter ground state, the heavy meson fields whose parameters scale
as (\ref{BR}) are way off-shell. In matter-free space their masses are comparable to the chiral
scale $\Lambda_\chi$ and hence one might naively think
that processes involving excitations of such particles on-shell
could not be handled reliably by the argument based on chiral symmetry used here.
Now what is observed in the CERES experiment
is highly excited modes, involving hundreds of MeV. In particular the
``$\rho$ meson'' which plays an important role in the description of
~\cite{LKB}
is near its mass shell {\it albeit at a scaled mass} and moving in the medium with certain momentum.
Thus it may be puzzling that the quasiparticle picture for the mesons works
so well. One would have expected that even within the given scheme, higher loop
graphs (e.g., widths) and explicit momentum dependences should enter importantly. This puzzle is
further highlighted by the equally
successful explanation of the same process by a description that
is based on standard many-body approach starting from a theory defined at zero density~\cite{wambach} which relies on the mechanism that in medium,
the width of the $\rho$ meson
increases~\cite{KW} or the $\rho$ ``melts."
\subsection{Approximate duality?}

One possible solution to the above puzzle is that as alluded already,
the description based on BR-scaling chiral
Lagrangians and the one based on many-body hadronic interactions are ``dual" in the sense that
they represent the same physics~\footnote{The duality I am referring to here
resembles in some sense the ``quark-hadron" duality much discussed in the
literature~\cite{duality}.}.
What the CERES data are telling us is that this duality
may be holding in the heat bath and that the two descriptions may be mapped
to each other~\cite{geb}. This may be understood in terms of a
``mended symmetry"~\cite{mended}.
As interpreted in \cite{klr}, the mended symmetry argument goes
as
follows. While in matter-free space, chiral symmetry is non-linearly
realized with
the massive scalar degree of freedom purged from the low-energy sector, as density
increases, a scalar, say, the $\sigma$, comes down to join the triplet of the pion
to ``mend" the
$O(4)$ symmetry of the chiral $SU(2)\times SU(2)$ and to become the fourth component of the
four vector. That is, in dense medium, the non-linear $\sigma$ model is
``mended" to
the linear $\sigma$ model with the masses BR-scaling~\cite{klr}.
How this can happen in nuclear dynamics with the broad scalar in
matter-free space becoming a local field in dense medium is described in \cite{brPR}. Now the
vector mesons, as long as they are still heavy, can be introduced much like the nucleon
as matter fields with their
masses scaling as (\ref{BR}). As density increases further and approaches
the critical density for the chiral transition, then the vector mesons become light and the Georgi
vector symmetry~\cite{georgi,brPR} would be ``mended."
{\it This interpretation clearly puts more significance on the
symmetry consideration
than on the complex dynamics (e.g., used in \cite{wambach}), in
conformity with what has been established in QCD at
long wavelength, namely, that in low-energy regime,
it is chiral symmetry of QCD that
governs the
physics of hadrons}. I suspect that it is this aspect that is at the root of the duality
we see in the CERES data. It would be extremely interesting to see
whether this dual description continues to hold
true when heavy mesons are probed in cold nuclear matter
as in Jefferson Laboratory or denser (somewhat warm) matter as
in HADES.

\section*{Acknowledgments}

The ideas presented in this article are largely based on work done over the
years in
collaboration with Gerry Brown, Bengt Friman, Kuniharu Kubodera, Chang-Hwan
Lee, Dong-Pil Min, Tae-Sun Park and Chaejun Song, all of whom I would like
to thank for stimulating discussions. I would also like to acknowledge
the generous support of my home institute, Service de Physique Th\'eorique
of Saclay, and the
hospitality of the theory groups at Stony Brook, GSI, TU M\"unchen and
the University of Valencia, and the Center for
Theoretical Physics of Seoul National University. Some of the more recent
work were supported by a Franco-German Humboldt Research Prize (GSI and Munich)
and by the IBERDROLA Visiting Professorship (Valencia).



\begin{thebibliography}{99}
\bibitem{aps} M. Rho, ``Chiral symmetry and meson-exchange currents,''
nucl-th/9708060
\bibitem{weinnature} S. Weinberg, Nature {\bf 386}, 234 (20 March 1997)
\bibitem{austern} N. Austern, \pr\ {\bf 92}, 670 (1953)
\bibitem{riskabrown} D.O. Riska and G.E. Brown, \pl\ {\bf B38}, 193 (1972)
\bibitem{pmr} T.-S. Park, D.-P. Min and M. Rho, \prl\ {\bf 74}, 4153 (1995);
\np\ {\bf A596}, 515 (1996)
\bibitem{kdr} K. Kubodera, J. Delorme and M. Rho, \prl\ {\bf 40}, 755 (1978).
For a prelude to this idea, see M. Chemtob and M. Rho, \np \ {\bf A163}, 1
(1971)
\bibitem{mr91} M. Rho, \prl\ {\bf 66}, 1275 (1991)
\bibitem{PKMR} T.-S. Park, K. Kubodera, D.-P. Min and M. Rho,
``Effective field theory for two-nucleon systems," to appear
\bibitem{v18} R.V. Wiringa, V.G.J. Stoks and R. Schiavilla, \pr\ {\bf C 51},
38 (1995)
\bibitem{lepage} G.P. Lepage, ``How to Renormalize the Schr\"odinger Equation,"
Lectures at the VIII Jorge Andr\'e Swieca Summer School, Brazil, February 1997;
nucl-th/9706029
\bibitem{frois} See, e.g., B. Frois and J.-F. Mathiot, Comments Part. Nucl.
Phys. {\bf 18}, 291 (1989)
\bibitem{warburton} E.K. Warburton, \prl\ {\bf 66}, 1823 (1991); \pr
\ {\bf C44}, 233 (1991); E.K. Warburton and I.S. Towner,
\pl\ {\bf B294}, 1 (1992)
\bibitem{CND} For a systematic account of this development, see
{\it Chiral Nuclear Dynamics}\ (World Scientific, Singapore, 1996) by
M.A. Nowak, M. Rho and I. Zahed
\bibitem{BRscaling} G.E. Brown and M. Rho, \prl\ {\bf 66}, 2720 (1991)
\bibitem{lynn} B.W. Lynn, \np\ {\bf B402}, 281 (1993)
\bibitem{SBMR} Chaejun Song, G.E. Brown, D.-P. Min and M. Rho, \pr {\bf C56},
2244 (1997), hep-ph/9705255
\bibitem{pmraxial} T.-S. Park, D.-P. Min and M. Rho, Physics Reports {\bf 233},
341 (1993);
T.-S. Park, I.S. Towner and K. Kubodera, \np\ {\bf A579}, 381 (1994)
\bibitem{jin} X. Jin and D.B. Leinweber, \pr\ {\bf C52}, 3344 (1995)
\bibitem{riska} M. Kirschbach, D.O. Riska and K. Tsushima, \np\ {\bf A542},
14 (1992); I.S. Towner, \np\ {\bf A542}, 631 (1992)
\bibitem{llb97} G.Q. Li, C.-H. Lee and G.E. Brown, nucl-th/9706057
\bibitem{LLB2}  G.Q. Li, C.-H. Lee and G.E. Brown, ``Kaon production
in heavy-ion collisions and maximum mass of neutron stars,''  nucl-th/9711002
\bibitem{brownbethe} G.E. Brown and H.A. Bethe, Astrophys. J. {\bf 423}, 659
(1994)
\bibitem{migdal} A.B. Migdal, {\it Theory of Finite Fermi Systems and
Applications to Atomic Nuclei} (Interscience Publishers, New York, 1967)
\bibitem{matsui} T. Matsui, \np\ {\bf A370}, 365 (1981)
\bibitem{shankar}
J. Polchinski, {\it Recent Directions in Particle Theory: From
Superstrings and Black Holes to the Standard Model}\ (TASI-92),
eds. J. Harvey and J. Polchinski (World Scientific, Singapore,
1994); R. Shankar, Rev. Mod. Phys. {\bf 66}, 129 (1994); T. Chen,
J. Fr\"olich and M. Seifert, cond-mat/9508063
\bibitem{br96} G.E. Brown and M. Rho, \np\ {\bf A596}, 503 (1996)
\bibitem{mnsr} R.J. Furnstahl, X. Jin and D.B. Leinweber, \pl\ {\bf B387},
253 (1996)
\bibitem{FR95} B. Friman and M. Rho, \np\ {\bf A606}, 303 (1996)
\bibitem{schumacher} R. Nolte, A. Baumann, K.W. Rose and M. Schumacher, \pl
\ {\bf B173}, 388 (1986)
\bibitem{yamazaki} T. Yamazaki, in {\it Mesons in Nuclei}, eds. M. Rho and
D.H. Wilkinson (North-Holland, Amsterdam, 1979)
\bibitem{gaexp} D.H. Wilkinson in {\it Physics with Heavy Ions and Mesons},
eds. R. Balian, M. Rho and G. Ripka (North-Holland, Amsterdam,
1978); B.H. Wildenthal, M.S. Curtin and B.A. Brown, \pr\ {\bf C28},
1343 (1983); B. Buck and S.M. Perez, \prl\ {\bf 50}, 1975 (1983);
K. Langanke et al, nucl-th/9504019
\bibitem{ROW} M. Rho, \np\ {\bf A231}, 493 (1974); K. Ohta and M. Wakamatsu,
\np\ {\bf A234}, 445 (1974)
\bibitem{PJM} T.-S. Park, H. Jung and D.-P. Min, nucl-th/9704033
\bibitem{furnstahl} R.J. Furnstahl, S.D. Serot and H.-B. Tang, \np\
{\bf A615}, 441 (1997)
\bibitem{SMR} Chaejun Song, D.-P. Min and M. Rho, ``Thermodynamic properties of
effective chiral Lagrangians with BR scaling,'' to appear
\bibitem{baymchin} G. Baym and S. Chin, \np \ {\bf A262}, 527 (1976)
\bibitem{ceres} G. Agakichiev et al, \prl\ {\bf 75}, 1272 (1995)
\bibitem{LKB} G.Q. Li, C.M. Ko and G.E. Brown, \prl\ {\bf 75}, 4007 (1995)
\bibitem{wambach} W. Cassing, E.L. Bratkovskaya, R. Rapp and J. Wambach,
nucl-th/9708020
\bibitem{KW} F. Klingl and W. Weise, \np\ {\bf A606}, 329 (1996)
\bibitem{duality} B. Grinstein and R.F. Lebed, ``Explicit quark-hadron
duality in heavy-light meson weak decays in the 't Hooft model,"
hep-ph/9708396
\bibitem{geb} G.E. Brown, private discussion
\bibitem{mended} S. Weinberg, \prl\ {\bf 65}, 1177 (1990)
\bibitem{klr} Y. Kim, H.K. Lee and M. Rho, \pr\ {\bf C5}, R1184 (1995); Y. Kim
and H.K. Lee, \pr {\bf C55}, 3100 (1997), nucl-th/9609020
\bibitem{brPR} G.E. Brown and M. Rho, Phys. Repts. {\bf 269}, 333 (1996)
\bibitem{georgi} H. Georgi, \np\ {\bf B331}, 311 (1990)
\end{thebibliography}
\end{document}